\documentclass[draftcls,12pt,onecolumn,oneside]{IEEEtran}
\usepackage{mathrsfs}
\usepackage{algorithm}
\usepackage{algorithmic}
\usepackage{amsthm}
\usepackage{acronym}
\usepackage{mathrsfs}
\usepackage{ifthen}
\usepackage{textcomp}
\usepackage{float}
\usepackage{subfigure}
\usepackage{color}
\usepackage{graphicx, epstopdf}
\usepackage{amssymb, amsmath}
\usepackage{commath, cite}
\usepackage{setspace}
\usepackage{latexsym} %
\usepackage{url,times,amsmath,color,amssymb,graphicx,epsfig,psfig,cite,geometry,psfrag,subfigure,dsfont,booktabs, multirow}
\begin{document}
\title{Network Slicing Based 5G and Future Mobile Networks: Mobility, Resource Management, and Challenges}
\author{Haijun Zhang,~\IEEEmembership{Member,~IEEE}, Na Liu, Xiaoli Chu,~\IEEEmembership{Senior Member,~IEEE}, \\
  Keping Long,~\IEEEmembership{Senior Member,~IEEE}, Abdol-Hamid Aghvami,~\IEEEmembership{Fellow,~IEEE}, \\ and Victor C.M. Leung,~\IEEEmembership{Fellow,~IEEE}
\thanks{Haijun Zhang and Keping Long are  with the Beijing Engineering and Technology Research Center for Convergence Networks and Ubiquitous Services, University of Science and Technology Beijing, Beijing, 100083, China. (e-mail: dr.haijun.zhang@ieee.org, longkeping@ustb.edu.cn).

Victor C.M. Leung is with the Department of Electrical and Computer Engineering, The University of British Columbia, Vancouver, BC V6T 1Z4 Canada (e-mail: vleung@ece.ubc.ca).

Na Liu is with College of Information Science and Technology, Beijing University of Chemical Technology, Beijing, 100029, China (Email: eeliuna@gmail.com).

Xiaoli Chu is with Department of Electronic and Electrical Engineering, the University of Sheffield, Sheffield S1 3JD, UK (Email: x.chu@sheffield.ac.uk).

A.-H. Aghvami is with the Department of Informatics, King's College London, London WC2R 2LS, U.K. (e-mail: hamid.aghvami@kcl.ac.uk).

}}\maketitle

\begin{abstract}
The fifth-generation (5G) networks are expected to be able to satisfy users' different quality-of-service (QoS) requirements. Network slicing is a promising technology for 5G networks to provide services tailored for users' specific QoS demands. Driven by the increased massive wireless data traffic from different application scenarios, efficient resource allocation schemes should be exploited to improve the flexibility of network resource allocation and capacity of 5G networks based on network slicing. Due to the diversity of 5G application scenarios, new mobility management schemes are greatly needed to guarantee seamless handover in network slicing based 5G systems. In this article, we introduce a logical architecture for network slicing based 5G systems, and present a scheme for managing mobility between different access networks, as well as a joint power and subchannel allocation scheme in spectrum-sharing two-tier systems based on network slicing, where both the co-tier interference and cross-tier interference are taken into account. Simulation results demonstrate that the proposed resource allocation scheme can flexibly allocate network resources between different slices in 5G systems. Finally, several open issues and challenges in network slicing based 5G networks are discussed, including network reconstruction, network slicing management and cooperation with other 5G technologies.


\end{abstract}

\section{Introduction}
With the rapid development and innovations of mobile networking technologies, an entirely new era of mobile communications, i.e., the fifth generation (5G) of mobile communication systems, is coming. There is a consensus that 5G systems can be rolled out around 2020. 5G systems are expected to provide the society with full connection, which can break through the limitations of time and space to create all-dimensional, user-centered or service-centric interconnections between people and things\cite{architecture1}.

5G networks aim to meet various user quality-of-service (QoS) requirements in different application scenarios, e.g., in terms of data transmission rate and latency\cite{scenarios2}. In scenarios where seamless wide-area coverage is needed, 5G systems should provide users with seamless high-data-rate services anytime and anywhere, even at cell edges or with high-speed (up to 500 km/h) mobility. In metropolitan areas where the density and volume of wireless traffic demand are both very high, 5G networks should provide dense hot-spot coverage with high capacity. In scenarios where reliable connections of a large number of widespread low-power nodes, e.g., wireless sensors, are needed, 5G networks should be able to connect millions of devices under the constraints of low power consumption and low cost per device. Extremely low latency and high reliability of 5G networks are required to meet the performance requirements of real-time, reliable and secure communications in some vertical industries such as interconnected vehicles and industrial production control.

Faced with the abundant, distinct, customized service requirements and in new application scenarios as mentioned above, the network architecture and networking technologies need to be revisited for 5G systems\cite{paradigm3}. This has become the focus of research and development activities of operators, equipment vendors and research institutes all over the world. In order to provide customized reliable services using limited network resources while reducing capital expenditure and operating expense of 5G networks, network slicing has recently been proposed by the wireless industry as a main enabler of network service convergence and on-demand customized services\cite{architecture4, slicing5, whitepaper6}. By slicing a physical network into several logical networks, network slicing can support on-demand tailored services for distinct application scenarios at the same time using the same physical network. Supported by network slicing, network resources can be dynamically and efficiently allocated to logical network slices according to the corresponding QoS demands.

Network slicing has also attracted a lot of research interests from the academia. In\cite{SDN7}, user-centric service slicing strategy considering different QoS requirements was proposed based on software defined networking (SDN), and a genetic algorithm was devised to optimize the virtualized radio resource management based on resource pooling. In\cite{mobility8}, a network slicing mechanism was introduced for network edge nodes to offer low-latency services to users, where the centralized core network (CN) entities and related applications are shifted to the network edge to reduce delays and burdens on the backhaul. The authors\cite{mobility8} also proposed mobility management schemes and an optimal gateway selection algorithm to support seamless handover. A resource allocation scheme with the consideration of interference management was presented in\cite{resource9}, where heterogeneous QoS requirements were guaranteed by optimizing power and subchannel allocation jointly. In\cite{mobility10}, an agile and flexible SDN based 5G network architecture was proposed to allocate physical network resources to virtual slices within a local area and to perform scheduling among slices. The SDN based network architecture features an unified control plane, where hierarchical controllers are used to achieve differentiated services in user access layers close to the base stations, radio access network (RAN) and CN, respectively. The research on mobility management in network slicing systems has mainly been focused on SDN based control and handover procedures\cite{mobility10,handover11,handover12}. In the existing literature, mobility management and virtualized resource allocation have not been sufficiently studied for network slicing based 5G networks.

In this article, we present a logical architecture for network slicing based 5G systems, including an introduction of the fundamental concepts of network slicing. Based on the proposed network architecture, we investigate mobility management and virtualized radio resource allocation technologies in network slicing based 5G systems. Due to the diversity and complexity of 5G scenarios, it is vital to study proper mobility management for different mobility scenarios. Accordingly, we present a handover management scheme for handovers between different access networks. Virtualized resource management is responsible for inter-slice and intra-slice allocation of network resources in a dynamic and efficient manner. We propose a joint power and subchannel allocation scheme for network slicing based spectrum-sharing two-tier networks, where both the co-tier interference and cross-tier interference are taken into account. Simulation results will show that the proposed resource allocation scheme can flexibly allocate network resources between different slices, thereby realizing the efficient sharing of network resources in 5G systems. Finally, we highlight the future challenges and open issues on network slicing in 5G systems.

The remainder of this paper is arranged as follows. The network slicing based 5G networks architecture is given, and network slicing management is described in detail in Section II. Next, in Section III, mobility management in 5G networks based on network slicing is simply discussed. In Section IV, a joint power and subchannel allocation scheme in spectrum-sharing two-tier systems based on network slicing is formulated and the simulation results are given.  Several open issues and challenges in network slicing based 5G networks are discussed in Section V. Lastly, the paper is summarized in VI.

\section{Network Slicing Based 5G System Architecture}
\begin{figure}[htbp]
        \centering
        \includegraphics*[width=15cm]{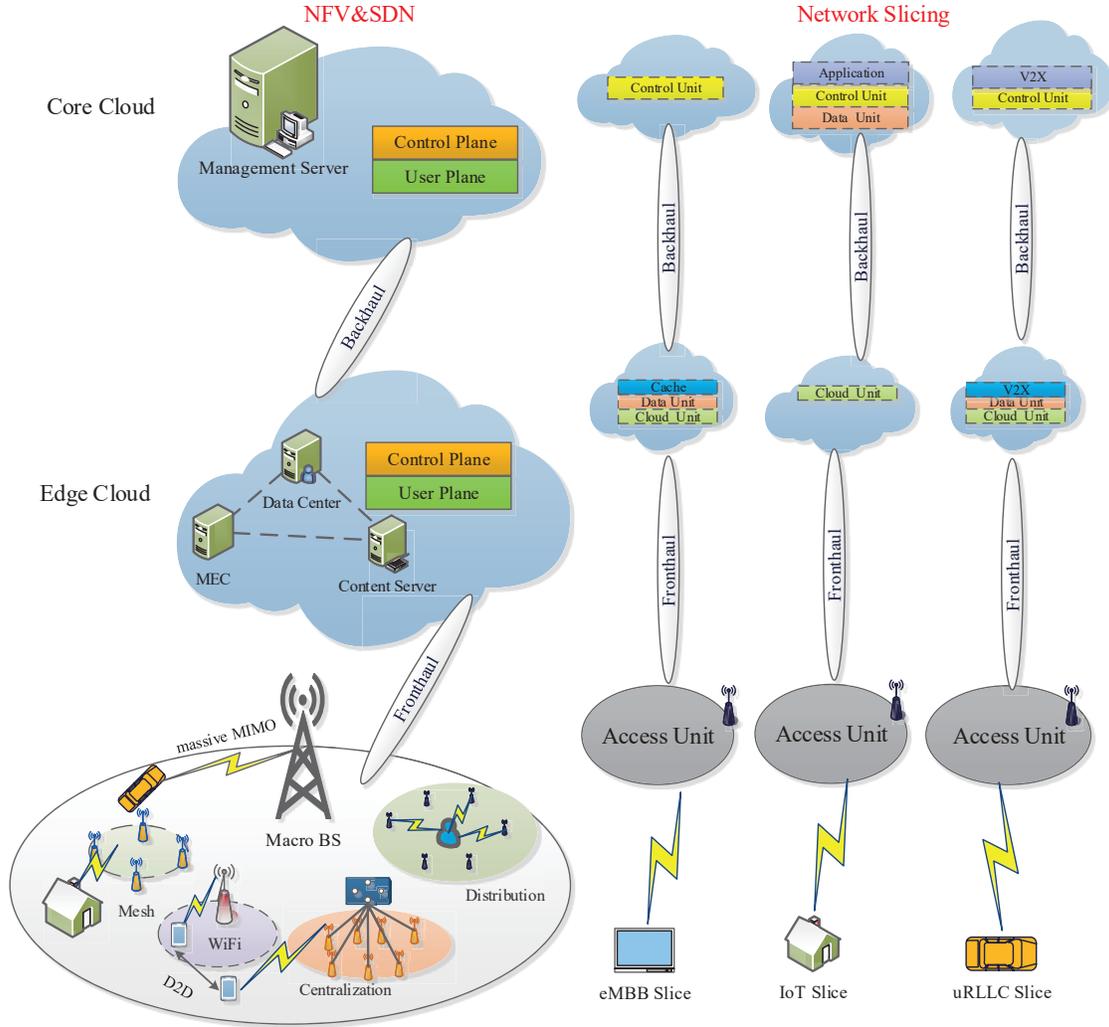}
        \caption{Network slicing based 5G system architecture.}
        \label{architecture}
\end{figure}

The design of 5G network architecture should be based on a comprehensive consideration of software control and hardware infrastructure and the interworking between them. Network slicing, which can fulfil diversity of network requirements based on the unified physical infrastructure and sharing network resources, is considered as a key paradigm to provide several independently operating instances for a specific network function\cite{slicing5}. SDN has been widely accepted as a promising technique to implement network slicing on the basis of network function virtualization (NFV)\cite{mobility10}. NFV replaces the traditional network elements (such as mobility management entity (MME), policy and charging rules function (PCRF), packet/service-gateway (P/S-GW)) in the CN and the RAN with commercial off-the-shelf servers, which also host the functions of dedicated physical infrastructures. Each such server can be considered as a pool of virtual machines (VM) running on commercial off-the-shelf hardware and software. The traditional RAN is divided into centralized processing units (e.g., baseband units (BBU) in cloud RAN (C-RAN)) and radio access units. The centralized processing units are largely virtualized, where resource pooling is introduced to perform service slicing in accordance with different QoS requirements\cite{challenges13}.

The logical architecture of a 5G system based on network slicing is given in Fig. 1. In the radio access plane of the 5G system, a heterogeneous network accommodates multiple radio access technologies (RATs) and supports the efficient cooperation between them. Small cells and WiFi access points are densely deployed to meet the increasing data traffic demand in 5G systems\cite{resource14}. Furthermore, device-to-device (D2D) communications are used to increase system capacity and improve energy- and spectrum-efficiency while reducing communication delays and relieving backhaul burden of macrocells\cite{mobility10}. D2D communications will play a critical role in network slicing based 5G systems, especially for improving quality of local services, emergency communications and Internet of things (IoT).

As shown in Fig. 1, the traditional centralized architecture of the CN has evolved into a core cloud, which separates the control plane from the user plane so as to reduce control signaling and delays of data transmissions. The core cloud provides some important functions of the control plane, such as mobility management, virtualized resource management, interference management, and so on. The servers and other functions of the RAN are located in the edge cloud, which is a centralized pool of virtualized functionalities. The edge cloud mainly performs data forwarding and control plane functions such as baseband processing. The user-plane functions in P/S-GW are also shifted to the edge cloud, to provide low-latency services and to reduce the burden on the backhaul. Mobile edge computing platforms are also deployed in the edge cloud, in conjunction with data forwarding and content storage servers, which can collaboratively execute the storage, computing, transmission of massive data in a real-time and efficient way. The corresponding VMs will distribute in core cloud and edge cloud to execute virtualized network functionalities. By utilize SDN, 5G networks can connect the VMs distributed in core cloud and edge cloud, creating the mapping between core cloud and edge cloud. Furthermore, the SDN controllers can control network slicing in a centralized fashion.
\begin{figure}[htbp]
        \centering
        \includegraphics*[width=15cm]{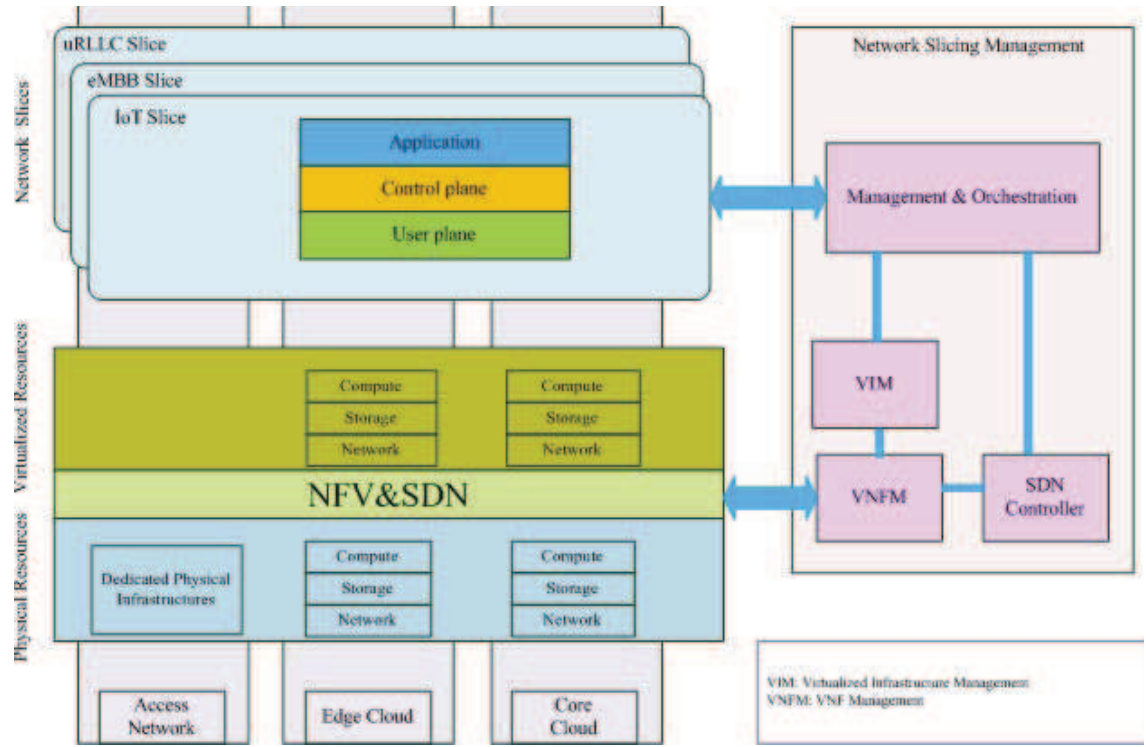}
        \caption{Network slicing management.}
        \label{slicing}
\end{figure}

After the virtualization and software redefinition of system architecture as described above, network slicing can be implemented. An example of network slicing operating on a set of generic physical infrastructures is illustrated in Fig. 2. An end-to-end network slice is a specific collection of network functions and resource allocation modules isolated from other network slices\cite{slicing5}. For example, the enhanced mobile broadband (eMBB) slice requires a large bandwidth to support high-data-rate services, such as high-definition video streaming and augmented reality. Caching function, data unit and cloud unit are also needed to assist control functions in implementing eMMB slicing services. Reliability, low-latency and security will be critical for the ultra-reliable and low-latency communication (uRLLC) slice to provide services that are extremely sensitive to latency, such as autonomous driving and Internet of vehicles (V2X). For uRLLC slice, all dedicated functions should be instantiated at the edge cloud. For IoT slice that serves a large number of static or dynamic machine type devices (such as sensors and monitors), the vertical applications will be placed on the upper layer to support the external services demanded by different commercial tenants.

In network slicing management, the control parts interact with each other through controllers or some kind of interfaces. The virtualized network function manager is responsible for the mapping of physical network functions to VMs. Coordinated with virtualized network function management (VNFM), the SDN controller operates and controls the entire virtual network by connecting the data layer and vertical applications through the interface protocols. Virtualized infrastructure management (VIM), as the center of the virtualized infrastructure, allocates virtualized resources to VMs by monitoring their resource utilization status. The network management and orchestration unit is the core part of slicing management, because it is responsible for creating, activating or deleting network slices according to customized service requirements.

The network slicing based 5G network architecture will radically change the traditional network planning and deployment patterns. Network slicing is driven by and tailored for the network applications and user requirements. By avoiding mapping each application to a single pipeline in the physical network, 5G networks can provide end-to-end tailored services according to customized application requirements.

\section{Mobility Management in Network Slicing based 5G Systems}
\begin{figure}[htbp]
        \centering
        \includegraphics*[width=15cm]{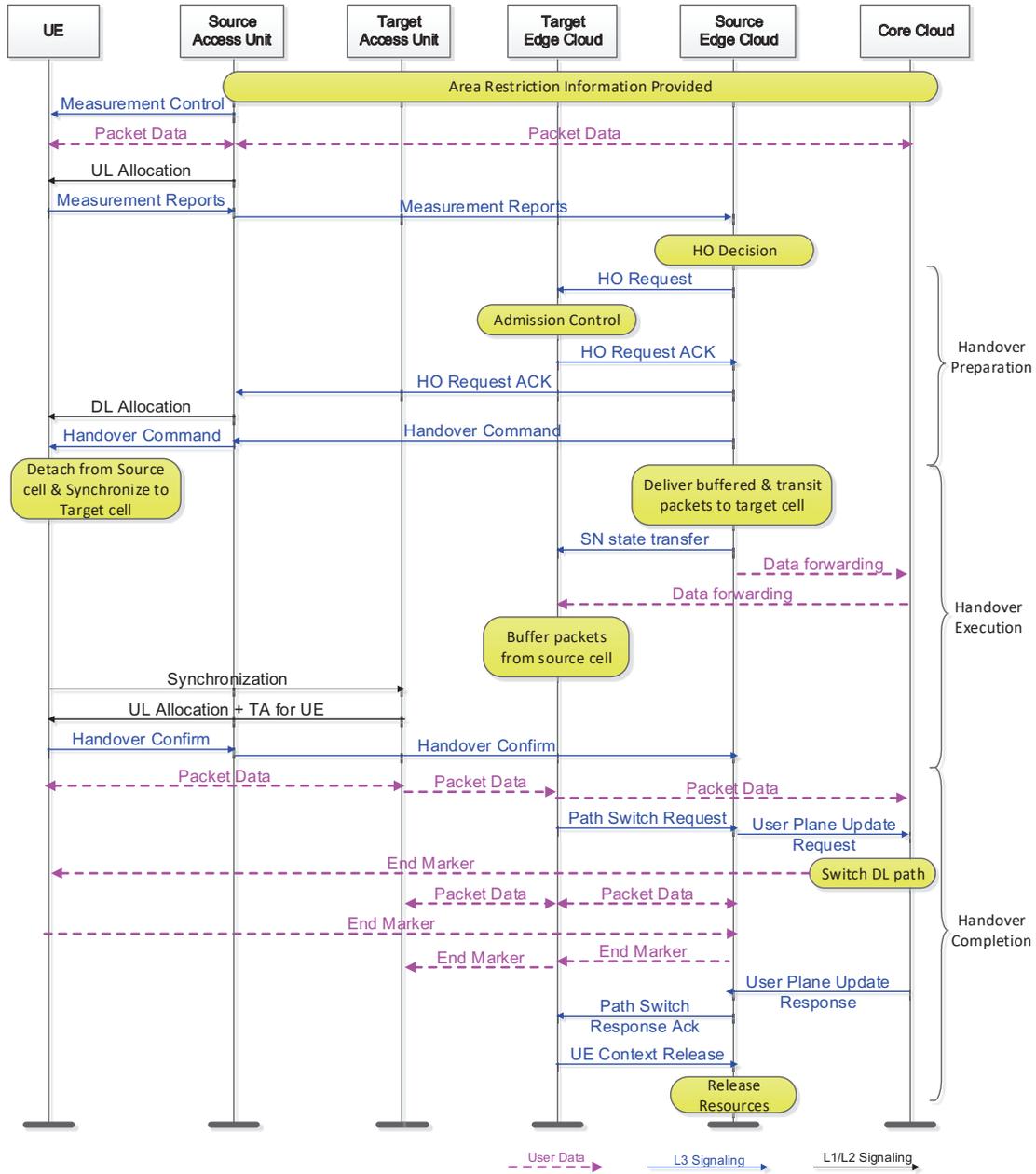}
        \caption{Handover procedure based on 5G network slicing systems.}
        \label{HandoverFlow}
\end{figure}

Mobility management in mobile communications has evolved from handling simple and single-RAT handover cases to managing complex, multi-RAT mobility scenarios. Based on SDN, the control plane and the user plane are split and decoupled at the gateway in the CN, and the integrated control functions can reduce control signaling even for a large number of distributed network nodes. However, network slicing based 5G systems will still face mobility management challenges caused by the potentially ultra high density of 5G networks combined with high mobility and high density of end devices. Consequently, new mobility management schemes need to be developed for network slicing based 5G systems to support seamless user experience with quality, continuity and scalability \cite{handover12}.

Different network slices have different characteristics and requirements in terms of mobility, latency and reliability. For instance, in railway communications, many handovers could be triggered by a high-speed train during a short time\cite{handover11}; while in IoT applications, reliable and/or low-latency communications should be guaranteed for many devices with low or no mobility. In the following, we study the on-demand and scalable mobility management mechanism under network slicing for customized service scenarios.

There are two main procedures in mobility management: location registration and handover management.

\subsection{Location registration}
Mobile devices register their locations when they first connect to the network, and then report their location information to the network periodically. In 5G networks, the home subscriber servers will be distributed into the edge cloud, making them closer to end devices to shorten registration delays and reduce backhaul burdens. 5G networks will aggregate multiple heterogeneous RATs. To achieve unified multi-RAT access and seamless mobility in 5G networks, multi-RAT coordination is needed for different RATs to share location information of their mobile devices.

\subsection{Handover management}
In conventional cellular networks, handovers are mainly event-triggered. The base station controls the user terminals to execute the measurement and report the measured network status information to serving base station. However, in our proposed network slicing based 5G systems, mobility related events need to be redefined. For instance, handovers may occur in different slicing scenarios. Flexible handover mechanisms and adaptive handover thresholds should be exploited to support mobility management in service-tailored scenarios.

In the proposed mobility management scheme for network slicing based 5G systems, the SDN is introduced into the RAN, generating the software-defined wireless network (SDWN). In SDWN, the single or hierarchical control plane is deployed closed to the edge cloud to support centralized control plane handover decisions. One SDN controller can handle handovers in a single network slice. In a hierarchical control plane within SDN, it is necessary for controllers to cooperate\cite{mobility10}. A handover signaling procedure in network slicing based 5G systems is given in Fig. 3. The user supported by one of the slices is communicating with other terminals through core cloud when the handover is triggered. After handover is executed successfully, the data will be transmitted through target edge cloud and target access unit to the user from core cloud. Due to virtualization, physical network elements are replaced by corresponding logical servers in core cloud and edge cloud. Moreover, in order to simplify multi-RAT cooperation, only IP protocols are used to support signaling interactions in the control plane. Existing interfaces are made open so that an unified interface protocol can operate flexibly. The SDN controllers located in the core cloud, the edge cloud and the access plane cooperatively carry out handover management in complex application scenarios.

\section{Virtualized Resource Allocation with Interference Management}
Network slicing facilitates dynamic and efficient allocation of network resources to meet diverse QoS requirements\cite{slicing5}. In SDN and NFV enabled network slicing systems, network resources are virtualized and managed in the centralized resource pools\cite{SDN7}. Due to limited network resources and increasingly diversified network services, it is challenging to efficiently provision network resources to network slices with different QoS requirements. Moreover, the heterogeneous nature of 5G networks (e.g., different RATs, different cell sizes) also adds complexities to resource allocation\cite{resource9}. Especially for densely deployed spectrum-sharing small cells, efficient and flexible resource allocation schemes with interference awareness are needed\cite{interference15}.

In this section, we present a resource allocation scheme tailored for different QoS requirements of uRLLC slice, IoT slice and eMBB slice, which are the three fundamental categories of network slicing in 5G systems. For example in uRLLC slicing scenarios, communication devices are more sensitive to time delay and require lower transmission rate than those in other slices. There could be mutual interference between small cells and macrocells, which provide services (e.g., video streaming) for eMBB slice and for IoT slice, respectively.

\subsection{Modeling and formulation}
As shown in Fig. 1, the collocated small cells and macrocell compose a two-tier system in the radio access plane. Small cells receive two kinds of interference: cross-tier interference from the macrocell and co-tier interference from neighboring small cells. In this scenario, we model the uplink resource allocation problem as the maximization of uplink capacity on each subchannel for small cells considering the following constraints: 1) the maximum transmit power of each small cell user;
2) the minimum data rate requirement of each uRLLC user; 3) the threshold of total interference power received by the macrocell from small cell users;  4) a subchannel can be allocated to at most one user in each small cell during one transmission interval.

\subsection{Solution based on the Lagrangian dual decomposition method}
The above formulation results in a non-convex discrete objective function. By relaxing the binary subchannel allocation indictors into continuous real variables, we transform it into a convex continuous function, which can be solved using the Lagrangian dual decomposition method. To simplify the solution, we decompose the objective function into a master problem and \emph{K}$\times$\emph{N} sub-problems (for \emph{K} small cells and \emph{N} subchannels). The Karush-Kuhn-Tucker (KKT) conditions are used to get the optimal power allocation, and the sub-gradient method is exploited to update the Lagrangian multipliers to obtain the optimal subchannel allocation.

\subsection{Simulation results}
\begin{figure}[t]
    \centering
    \includegraphics[width=1.00\linewidth]{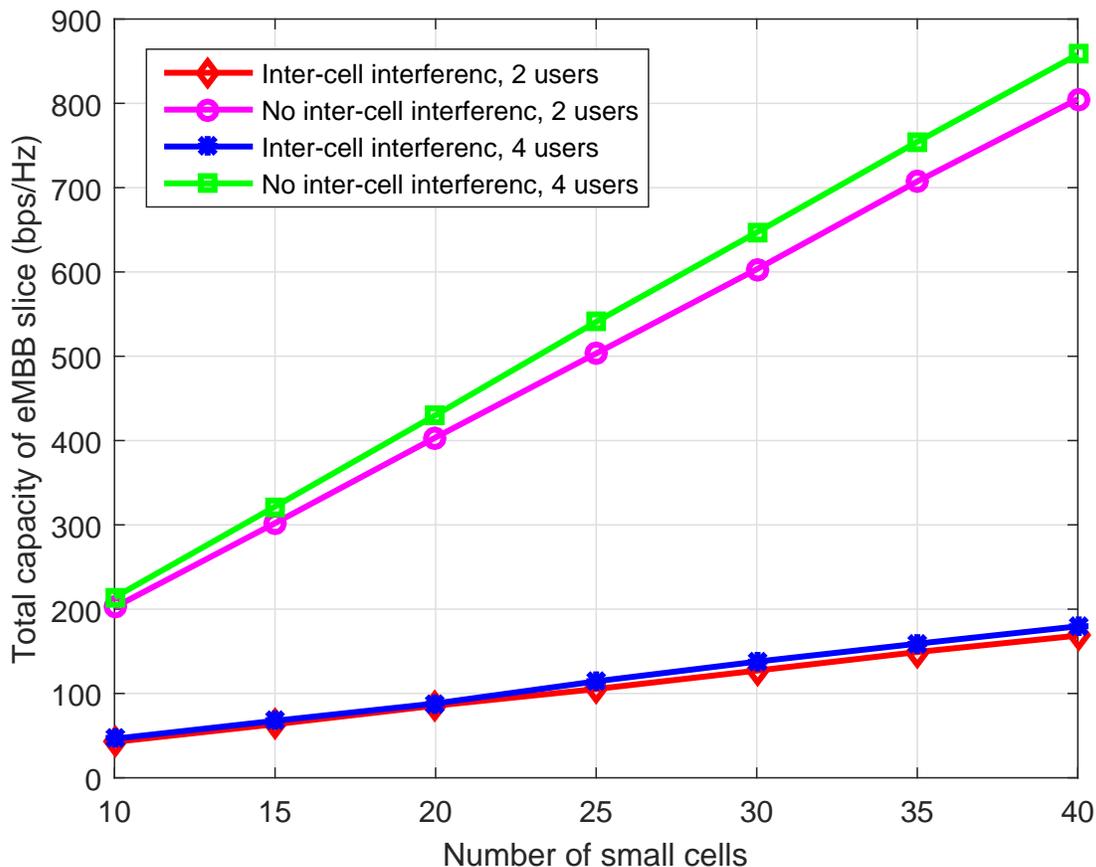}
    \caption{Total capacity of eMBB slice versus the number of small cells.}
    \label{eMBB}
\end{figure}

We present simulation results to demonstrate the performance of a network slicing based 5G network (in conjunction with the proposed subchannel and power allocation scheme), where a suburban environment is considered with small cells randomly distributed in the macrocell coverage area. The macrocell coverage radius is 500 meters and that of a small cell is 10 m. Other system parameters are set as follows: the carrier frequency is 2 GHz, the 10 MHz channel is divided into 50 subchannels, the minimum inter-small-cell distance is 20 meters, the maximal transmission power (of small cell and macrocell users) is 23dBm, the threshold of interference per subchannel (received by the macrocell) is -101.2 dBm, and the power spectral density of additive white Gaussian noise (AWGN) is -174 dBm/Hz. There are 50 users (requesting IoT services) distributed randomly in the macrocell, and 2 or 4 users (requesting uRLLC or eMBB services) camping on each small cell. The channel model includes path loss (indoor and outdoor) and frequency-selective fading. Round-robin scheduling is used in each cell, and uniform power allocation is adopted for macrocell users.
\begin{figure}[t]
    \centering
    \includegraphics[width=1.00\linewidth]{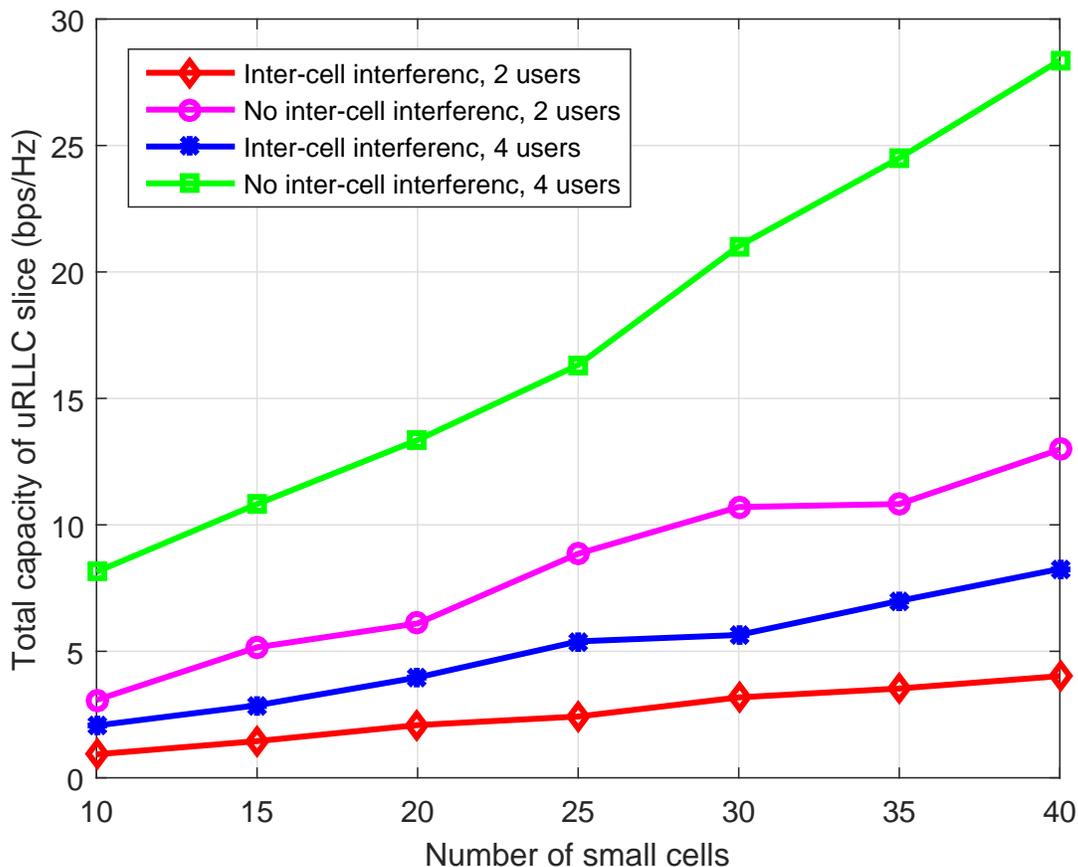}
    \caption{Total capacity of uRLLC slice versus the number of small cells.}
    \label{uRLLC}
\end{figure}

\begin{figure}[htbp]
    \centering
    \includegraphics[width=1.00\linewidth]{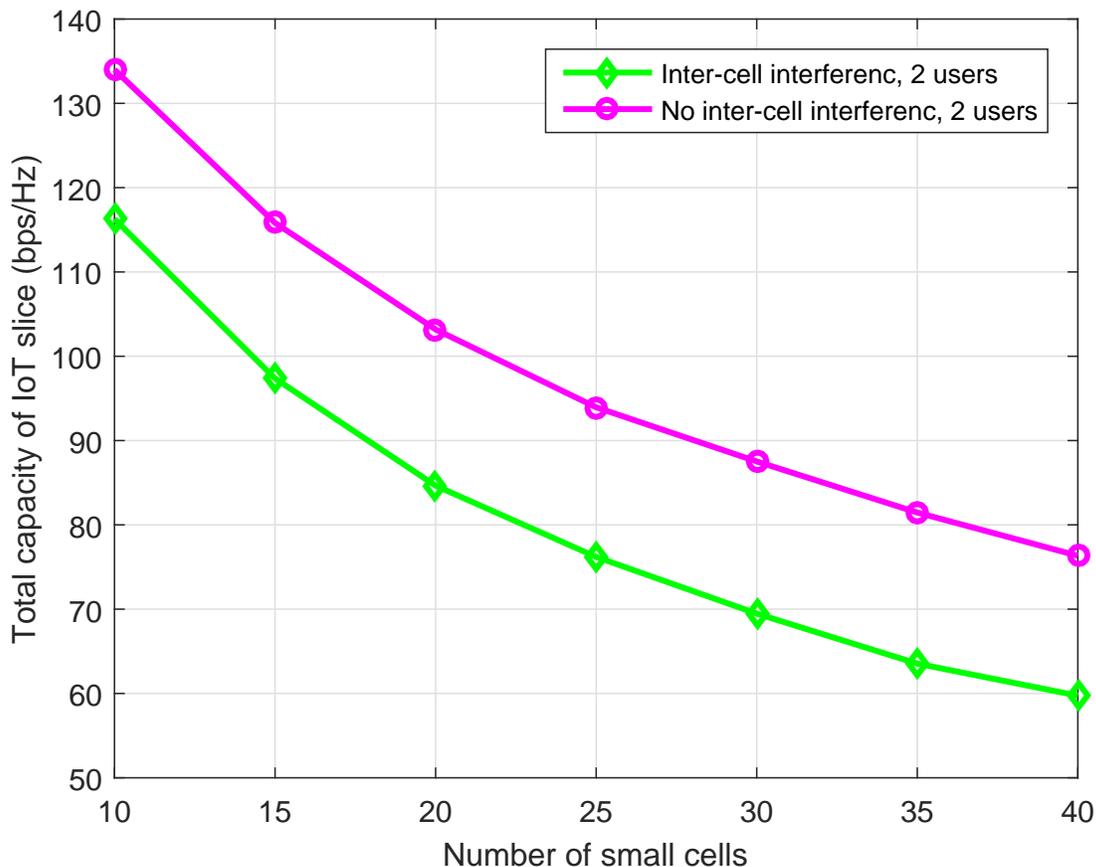}
    \caption{Total capacity of IoT slice versus the number of small cells.}
    \label{IoT}
\end{figure}

Fig. 4 shows the total capacity of the eMBB slice versus the number of small cells per macrocell. We can see that the eMBB slice capacity rises nearly linearly with the density of small cells and increases slightly with the number of users per small cell. However, the eMBB slice capacity decreases significantly due to the inter-cell interference between small cells, especially at high small cell densities.

Fig. 5 shows that the total capacity of uRLLC slice also increases with the number of small cells, but the capacity of uRLLC slice is 20 times less than that of eMBB slice. This is because the eMBB slice uses large bandwidths to transmit massive data, while the uRLLC slice only transmits low-volume control messages or data under low-latency constraints.

Fig. 6 shows the total capacity of the IoT slice supported by the macrocell, which suffers from cross-tier interference from small cells supporting the eMBB and uRLLC slices. The total capacity of the IoT slice decreases with the number of small cells, due to the increasing cross-tier interference caused. The capacity of IoT slice will further decrease due to co-tier interference between macrocells and between small cells. This is because with channel quality affected by increased inter-small-cell interference, small cell users will adaptively increase their transmit power leading to an increase of cross-tier interference from small cells.

The simulation results have shown that in both latency-sensitive and latency-tolerant network slicing scenarios, the proposed resource allocation scheme can allocate network resources properly and efficiently, and can improve system capacity of dense heterogeneous networks. Due to the space limitation, we will discuss other metrics (such as latency) in future works.

\section{Challenges and Open Issues}
Network slicing is a promising paradigm in future 5G mobile networks, but realizing it is not without challenges\cite{challenges13}. In the following, we discuss major challenges and open issues on network slicing in terms of network reconstruction, slicing management and cooperation with other 5G technologies.

\subsection{Network reconstruction}
Since 5G networks provide wireless connection for everything, both RAN and CN need reconstruction to support end-to-end network slicing. Especially in dense heterogeneous networks, not only the cooperation of macrocells and small cells should be designed to meet the customized slicing demands, but also the cooperation of multiple RATs should be considered to provide seamless mobility and high transmission throughput.

\subsection{Network slicing management}
Network slicing supports customized configuration of resources, management models, system parameters for various use cases in an isolated or abstract way\cite{slicing5}. Although service providers and mobile operators have started developing industrial solutions for network slicing, the management of network slicing is still a hard nut to crack. There are many dimensions and technologies included in network slicing: to create, activate, maintain and deactivate network slicing at the service level; to adjust load balance, charging policies, security and QoS at the network level; to abstract and isolate virtualized network resources; and inter-slice and intra-slice resource sharing. Moreover, the complexity and difficulty of network slicing management may increase with the continue booming of applications and services.

\subsection{Cooperation with other 5G technologies}
In future 5G systems, network slicing needs to coexist and cooperate with traditional technologies (such as broadband transmission, mobile cloud engineering (MCE), SDN and NFV) evolved from LTE/LTE-A systems. The virtualized cloud of access networks and CN have the advantages of physical resource pooling, distribution of software architectures, centralization of management. However, there is still no proper approach to integrate network slicing with C-RAN, SDN and NFV to provide point-to-point connection between physical radio equipment and radio equipment controller. Cooperation between network slicing and  other 5G technologies is necessary to enable more network slices in future 5G networks.

\section{Conclusion}
In this article, we have presented a logical architecture for network slicing based 5G systems, and discussed the evolution of network architecture based on SDN and NFV technologies, as well as the implementation of network slicing. Based on the network slicing architecture, we revised handover procedures in mobility management, and discussed mobility management mechanisms to offer flexible and agile customized services in network slicing based 5G systems. Moreover, considering various network slicing scenarios, we introduced a resource allocation mechanism tailored for QoS requirements and interference constraints of uRLLC, eMBB and IoT service slices. The promising performance of network slicing based 5G networks has been demonstrated through computer simulations.

\section*{Acknowledgment}
This work was supported by the National Natural Science Foundation of China (Grant 61471025), the Open Research Fund of The State Key Laboratory of Integrated Services Networks, Xidian University (ISN17-02), and the Fundamental Research Funds for the Central Universities.

\begin{IEEEbiography}{Haijun Zhang} is currently a Full Professor in University of Science and Technology Beijing, China. He was a Postdoctoral Research Fellow in Department of Electrical and Computer Engineering, the University of British Columbia (UBC), Vancouver, Canada. He received his Ph.D. degree in Beijing University of Posts Telecommunications (BUPT). From 2011 to 2012, he visited Centre for Telecommunications Research, King's College London, London, UK, as a Visiting Research Associate. Prof. Zhang has published more than 80 papers and authored 2 books. He serves as Editor of Journal of Network and Computer Applications, Wireless Networks, Telecommunication Systems, and KSII Transactions on Internet and Information Systems, and serves/served as a leading Guest Editor for IEEE Communications Magazine, IEEE Transactions on Emerging Topics in Computing and ACM/Springer MONET. He serves/served as General Co-Chair of GameNets'16 and 5GWN'17, Symposium Chair of the GameNets'14, Track Chair of ScalCom2015, and Co-Chair of Workshop on 5G Ultra Dense Networks in ICC 2017.
\end{IEEEbiography}

\begin{IEEEbiography}{Na Liu} received the BS degree in electronic information engineering from Beijing University of Chemical Technology, Beijing, China, in 2016. She is currently pursuing the M.S. degree at the Laboratory of Wireless Communications and Networks from College of Information Science and Technology, Beijing University of Chemical Technology, Beijing, China. Her research interests include resource allocation, power control, energy efficiency in wireless communications, software-defined wireless networks, and visible light communications.
\end{IEEEbiography}

\begin{IEEEbiography}{Xiaoli Chu} is a Senior Lecturer in the Department of Electronic and Electrical Engineering at the University of Sheffield, UK. She received the B.Eng. degree from Xi¡¯an Jiao Tong University in 2001 and the Ph.D. degree from the Hong Kong University of Science and Technology in 2005. From 2005 to 2012, she was with the Centre for Telecommunications Research at King¡¯s College London. She has published over 100 peer-reviewed journal and conference papers. She is the lead editor/author of the book Heterogeneous Cellular Networks ¨C Theory, Simulation and Deployment, Cambridge University Press, May 2013. She is Editor for the IEEE Wireless Communications Letters and the IEEE Communications Letters. She was Co-Chair of Wireless Communications Symposium for the IEEE International Conference on Communications 2015, and Workshop Co-Chair for the IEEE International Conference on Green Computing and Communications 2013.
\end{IEEEbiography}

\begin{IEEEbiography}{Keping Long}[SM] received his M.S. and Ph.D. degrees at UESTC in 1995 and 1998, respectively. He worked as an associate professor at BUPT. From July 2001 to November 2002, he was a research fellow in the ARC Special Research Centre for Ultra Broadband Information Networks (CUBIN) at the University of Melbourne, Australia. He is now a professor and dean at School of Computer and Communication Engineering (CCE), USTB. He is a member of the Editorial Committee of Sciences in China Series Fand China Communications . He is also a TPC and ISC member for COIN, IEEE IWCN, ICON, and APOC, and Organizing Co-Chair of of IWCMC'06, TPC Chair of COIN'05/'08, and TPC Co-Chair of COIN'08/'10, He was award-ed the National Science Fund Award for Distinguished Young Scholars of China in 2007 and selected as the Chang Jiang Scholars Program Professor of China in 2008. His research interests are optical Internet technology, new generation network technology, wireless information networks, value-added service, and secure network technology. He has published over 200 papers, 20 keynotes, and invited talks.
\end{IEEEbiography}

\begin{IEEEbiography}{Hamid Aghvami} joined the academic staff at King¡¯s College London in 1984. In 1989 he was promoted to Reader, and in 1993 was promoted Professor in Telecommunications Engineering. He is/was the founder/Director of the Centre for Telecommunications Research at King's. Professor Aghvami has published over 580 technical journal and conference papers, filed over 30 patents. He was an Executive Advisor/Chairman/Managing Director of many Wireless Communications companies. Professor Aghvami was a member of the Board of Governors of the IEEE Communications Society in 2001-2003, was a Distinguished Lecturer of the IEEE Communications Society in 2004-2007, and has been member, Chairman, and Vice-Chairman of the technical programme and organising committees of a large number of international conferences. He is also founder of the International Symposium on Personal Indoor and Mobile Radio Communications (PIMRC), a major yearly conference attracting some 1,000 attendees. Professor Aghvami was awarded the IEEE Technical Committee on Personal Communications (TCPC) Recognition Award in 2005. Professor Aghvami is a Fellow of the Royal Academy of Engineering, Fellow of the IET, Fellow of the IEEE, and in 2009 was awarded a Fellowship of the Wireless World Research Forum.
\end{IEEEbiography}

\begin{IEEEbiography}{Victor C. M. Leung} (S'75, M'89, SM'97, F'03) is a Professor of Electrical and Computer Engineering and holder of the TELUS Mobility Research Chair at the University of British Columbia (UBC).  His research is in the areas of wireless networks and mobile systems. He has co-authored more than 900 technical papers in archival journals and refereed conference proceedings, several of which had won best-paper awards. Dr. Leung is a Fellow of the Royal Society of Canada, a Fellow of the Canadian Academy of Engineering and a Fellow of the Engineering Institute of Canada. He is serving on the editorial boards of IEEE JSAC-SGCN, IEEE Wireless Communications Letters, IEEE Access and several other journals. He has provided leadership to the technical program committees and organizing committees of numerous international conferences. Dr. Leung was the recipient of the 1977 APEBC Gold Medal, NSERC Postgraduate Scholarships from 1977-1981, a 2012 UBC Killam Research Prize, and an IEEE Vancouver Section Centennial Award.
\end{IEEEbiography}

\end{document}